\documentclass[pre,twoside,showpacs,twocolumn,floatfix]{revtex4}

\usepackage{amssymb}
\usepackage{amsfonts}
\usepackage{amsmath}
\usepackage[dvips]{graphicx}
\usepackage{enumerate}
\usepackage{latexsym}
\usepackage{soul}
\usepackage[dvips,usenames,dvipsnames]{color}
\begin{document}
\sloppy
\title{The effect of finite container size on granular jet formation}
\author{Stefan von Kann$^{1}$, Sylvain Joubaud$^{1,2}$, Gabriel A. Caballero-Robledo$^{1,3}$, Detlef Lohse$^{1}$, and Devaraj van der Meer$^{1}$}
\affiliation{$^{1}$Physics of Fluids group, University of Twente, P.O. Box 217, 7500 AE Enschede, The Netherlands\\
$^{2}$Laboratoire de Physique de l'\'Ecole Normale Sup\'erieure de Lyon, CNRS \& Universit\'e de Lyon, F-69364 Lyon, France\\
$^{3}$Centro de Investigaci\'{o}n en Materiales Avanzados S. C., Nuevo Le\'{o}n, Mexico.}
\date{\today}

\begin{abstract}
When an object is dropped into a bed of fine, loosely packed sand, a surprisingly energetic jet shoots out of the bed. In this work we study the effect that boundaries 
have on the granular jet formation. We did this by (i) decreasing the depth of the sand bed and (ii) reducing the container diameter to only a few ball diameters. These confinements change the behavior of the ball inside the bed, the void collapse, and the resulting jet height and shape. We map the parameter space of impact with Froude number, ambient pressure, and container dimensions 
 as parameters. From these results we 
propose a new explanation for the thick-thin structure of the jet reported by several groups [J.R.Royer \textit{et al}., Nature Phys. \textbf{1}, 164 (2005)], [G.Caballero \textit{et al}., Phys. Rev. Lett. \textbf{99}, 018001 (2007)] and [J.O. Marston, \textit{et al}., Physics of Fluids \textbf{20}, 023301 (2008)].
\end{abstract}


\maketitle

\section{Introduction}
Granular materials consist of discrete particles which interact mainly through contact forces. In large quantities they can behave like a solid, a liquid, or a gas but often behave differently from what would be expected of these phases~\cite{Jaeger1996}. A marked example is the impact of an object into a bed of sand. When dry air is blown through such a bed all contact forces between the individual particles are broken and after slowly turning off the air flow, the bed settles 
into an extremely loosely packed solid-like state. When a ball is dropped in such a bed, one observes a splash and a jet, strikingly similar to the ones that are seen when the same object is dropped into a liquid.

Research interest in this granular jet started when S.T. Thoroddsen and A.Q. Shen first reported this phenomenon in 2001~\cite{refShen01}, in a study with the objective to gain insight into the importance of surface tension on jetting in general and the properties of flowing granular materials. Since these results, several aspects of the formation of the granular jet have been studied. The influence of the impact velocity onto the jet height for impacts on a bed of very loose sand was investigated in~\cite{refLohsePRL04}. Using a pseudo two-dimensional setup, numerical simulations  and comparisons to water impact experiments, a model for the jet formation was proposed that is based on cavity collapse: The impacting ball creates a cavity in the sand bed which collapses due to the hydrostatic pressure in the sand and leads to two vertical jets. One jet is observable above the bed and the other one is going down into the bed~\cite{refLohsePRL04}. The series of events is concluded by a ``granular eruption'' at the surface of the sand which was attributed to the surfacing of an air bubble that is entrapped during the collapse.

The influence of the ambient pressure on the formation of a granular jet was first studied by Royer \textit{et al.}~\cite{refRoyerNP05}. They observed that at lower ambient pressures the jet reaches less high and also reported a puzzling thick-thin structure at lower pressures. Using X-ray radiographic measurements, they were able to look inside the bed and then proposed the following mechanism to explain this structure: the thick jet is caused by the compressed air in the cavity pushing up bed material, forming the thick part of the jet~\cite{refRoyerNP05,refRoyerPRL07,refRoyerPRE08}. The thin jet was attributed to the hydrostatic collapse as formulated in~\cite{refLohsePRL04}. Subsequently, the thick-thin structure was also observed at atmospheric pressure by increasing the ball size in the same container, which suggests --in contrast to the earlier explanation-- that the structure may be a boundary effect~\cite{refCaballeroPRL07}. Marston {\em et al} also found a thick-thin structure by decreasing the packing fraction, and they too found that this effect is more pronounced for a larger ball~\cite{refMarston08}. It is the exploration of the formation of this thick-thin structure that constitutes the main motivation for the work described in the current paper.

In parallel to the research concerning the formation of the granular jet, quite some effort was made 
to understand the motion of an object moving through a granular medium. Different drag force laws were proposed~\cite{refUeharaPRL03, refCiamarraPRL04,refLohseNature04,refDeBruyn04,refHouPRE05,refTsimring05,refKatsuragi07}, culminating in a model containing a hydrostatic term that linearly depends on the depth inside the bed and a dynamic term which is proportional to the square of the velocity of the object~\cite{refTsimring05,refKatsuragi07}. The influence of the ambient air pressure on this trajectory was investigated in~\cite{refCaballeroPRL07,refRoyerPRL07} where it was shown that the drag force reduces at high ambient pressure. Another important issue is the interaction between the impacting ball and the container boundaries. Nelson \textit{et al.} found that ``the presence of sidewall causes less penetration and an effective repulsion''~\cite{refDurian2008,refSequin08}.

In this paper, we present experiments in which the size of the container has been systematically reduced. We did this by (i) decreasing the depth of the sand bed (section~\ref{shallow_bed}) and (ii) reducing the container diameter to only a few ball diameters (section~\ref{boundaries}). We explore how these confinements change the behavior of the ball inside the bed, the void collapse, the resulting jet height and shape, and the presence of a granular eruption, which was only observed in part of the parameter space covered in this study. All of the observed phenomena are explained within the context of a simple hydrostatic collapse model~\cite{refLohsePRL04} together with a drag law for the trajectory of the ball inside the sand~\cite{refKatsuragi07}. Finally, we propose an explanation for the presence of an eruption and a new mechanism for the thick-thin structure reported by several groups mentioned above.

The paper is organized as follows: In Section~\ref{Theory} we start with the introduction of the drag law and the hydrostatic collapse model that lie at the heart of the analysis of this paper. Subsequently we discuss our experimental setup in Section~\ref{exp_setup} after which we present our results for impacts in confined settings. In Section~\ref{shallow_bed} we discuss the influence of the proximity of the bottom, after which we turn to the influence of the side walls in Section~\ref{boundaries}. Finally, in Section~\ref{TTshapes} we discuss the thick-thin structure and end with conclusions in Section~\ref{conclusions}.

\section{Drag law and hydrostatic collapse model \label{Theory}}

\begin{figure}[htp]
\includegraphics[width=\linewidth]{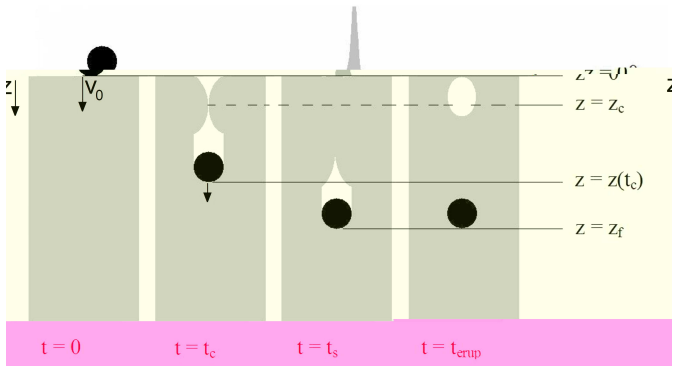}
\caption{Schematic representation of the impact of a ball into a sand bed, indicating the time and length scales that play an important role in the analysis of the experimental work in this paper, as described in the main text.} \label{schema_jet}
\end{figure}

In this Section we review the drag law and the Rayleigh-type collapse model that constitute the ingredients of the hydrostatic collapse model first introduced in \cite{refLohsePRL04} and form the theoretical backbone against which the experiments will be analyzed. Before doing so let us recall the succession of events after an impact of a sphere in a granular bed, presented in Fig.~\ref{schema_jet}, which involves the introduction of several time and length scales that are crucial to the analysis in the following Sections.  At a time $t\,=\,0$, the sphere impacts on the granular medium with a velocity $v_0$. A splash is created and the ball penetrates into the sand bed. The void created by the ball collapses in a time $t_c$ (closure time) and leads to the formation of two jets (one of them is visible above the bed and the other one is going down into the bed). The closure depth --also known as the pinch-off point-- is denoted as $z_c$ and the position of the ball inside the sand at that time as $z(t_c)$. At this time the ball will typically still be moving inside the sand bed, until, after a time $t_s$, the ball reaches its final depth $z_f$ and stops. Finally, the entrapped air bubble slowly rises inside the sand bed and leads to a granular eruption at the surface at $t=t_{\rm erup}$.

The first ingredient of the hydrostatic collapse model concerns the motion of the ball with diameter $d$ through the sand bed. To describe the trajectory of the ball ($z(t)$ is the depth of the ball at a time $t$), we use the law introduced by Tsimring~\cite{refTsimring05} and Katsuragi~\cite{refKatsuragi07}. The drag force is decomposed into two terms: The first one, the hydrostatic term, involves Coulomb friction as well as the force needed to displace material against the hydrostatic pressure and is proportional to the depth and was introduced in this context in~\cite{refLohseNature04}. We here write $F_{\rm hydrostatic} = \kappa z$ where $\kappa$ is a constant. The second term is an quadratic drag independent of the depth, $F_{\rm dynamic} = \alpha v^2$, interpreted as the quadratic force required for the projectile to mobilize a volume of granular media with density $\rho_g$ proportional to the volume of the ball~\cite{footnote1}. Adding gravity, this leads to the equation of motion:
\begin{equation}
m\ddot{z} =  mg - \kappa z - \alpha\,v^2\,,
\label{eq:balistic}
\end{equation}
with initial conditions $z(0) = 0$ and $\dot{z}(0) = v_0$.

The second ingredient regards the dynamics of the hydrostatic collapse of the void that is formed by the ball. The radius of the void at a time $t$ and a depth $z$, $R(z,t)$, evolves from the two-dimensional Rayleigh-type equation, in which, for each depth $z$, the collapse is driven by the hydrostatic pressure $\rho_g gz$ at that depth~\cite{refLohsePRL04}
\begin{equation}
(R\ddot{R}+\dot{R}^2)\log\frac{R}{R_{\infty}}+\frac{1}{2}\dot{R}^2 = g z\,,
\label{eq:Rayleigh}
\end{equation}
where $\dot{R}$ denotes the time derivative and $R_{\infty}$ is a constant of the order of the system size. Under the assumption that the cavity that is created is approximately cylindrical, i.e., with the same diameter ($d$) as the impacting ball, the initial conditions are $R(0)=d/2$ and $\dot{R}(0) = 0$. By rescaling lengths with the ball radius $d/2$ and time with $d/(2\sqrt{gz})$ (i.e., $\widetilde{R} \equiv 2R/d$, $\dot{\widetilde{R}} \equiv R/\sqrt{gz}$, etc., where the dot on a dimensionless variable denotes a derivative with respect to dimensionless time), Eq.~(\ref{eq:Rayleigh}) can be written in dimensionless form
\begin{equation}
(\widetilde{R}\ddot{\widetilde{R}}+\dot{\widetilde{R}}^2)\log\frac{\widetilde{R}}{\widetilde{R}_{\infty}}+\frac{1}{2}\dot{\widetilde{R}}^2 = 1\,,
\label{eq:RayleighDimless}
\end{equation}
together with initial conditions $\widetilde{R}(0) = 1$ and $\dot{\widetilde{R}}(0) = 0$. With these initial conditions this equation has a unique solution $\widetilde{R}(\widetilde{t})$, from which we obtain a constant dimensionless collapse time $\widetilde{t}_{\rm coll}$. It now follows immediately that the (dimensional) collapse time $t_{\rm coll}$ [$=\widetilde{t}_{\rm coll}d/(2\sqrt{gz})$] scales as $\sim d/(2\sqrt{gz})$.

Finally, we can combine the above two ingredients to determine the position and the time of closure. The total time that elapses from the impact to the collapse of the cavity at any depth $z$ is given by:
\begin{equation}
t_{\rm tot}(z) = t_{\rm pass}(z)+t_{\rm coll}(z)\,.
\label{eq:time}
\end{equation}
where $t_{\rm pass}$ is the amount of time the ball takes to reach depth $z$ (obtained from solving the drag law) and $t_{\rm coll}$ is the time needed for the collapse at a depth $z$. The closure depth is the depth which minimizes equation~\ref{eq:time}. The closure time corresponds to the total time at the closure depth ($t_c \equiv t_{\rm tot}(z_c)$).

\section{Experimental setup}\label{exp_setup}

\begin{figure}[htp]
\includegraphics[width=\linewidth]{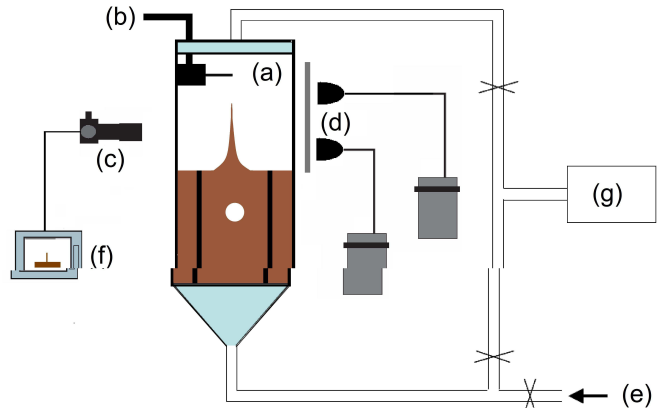}
\caption{Setup: (a) perspex container, $14 \times14 \times 100$ cm$^3$, (b) pneumatic release mechanism, (c) Photron Ultima APX-RS, (d) two light sources with diffusing plate, (e) pressurized, dry air source, (f) computer and (g) vacuum pump with pressure gauges.} \label{setup}
\end{figure}

In the previous section, we have introduced the theoretical framework for the analysis of the phenomenon. We now turn to the description of the experimental setup, which is sketched in Fig.~\ref{setup}. It consists of a container with a height of $1$~m and a square cross section of $14\,\times\,14$~cm$^2$, which is filled with sand grains, nonspherical and slightly polydisperse in size (between $20$ and $60$~$\mu$m); the density of the granular medium is $2.21$~g/cm$^3$ and its angle of repose $26^\circ$. As described in~\cite{refCaballeroPRL07}, before each experiment, the sand is fluidized by blowing pressurized dry air through a sintered plate at the container bottom. After slowly turning off the air flow, the bed reproducibly settles into a static, loose, weakened state (volume fraction $41$~$\%$). The airtight system can be slowly evacuated to perform experiments at lower ambient pressures $p$ (the pump speed is low enough not to irreversibly alter the packing fraction). Then a steel ball of diameter $d = 1.6$~cm and mass $m = 16.5$~g is dropped into the sand from different heights $H$ which controls the impact velocity $v_0 = \sqrt{2gH}$, where $g$ is the acceleration of gravity. Thus, the impactor is characterized by a single dimensionless number, the Froude number (Fr), defined as $\textrm{Fr} = 2v_0^2/(gd) = 4H/d$.

The impact is recorded with a high-speed camera (Photron Ultima APX-RS) at $1500$ frames per second. For the measurements a uniform lighting from behind is needed to obtain better movies with higher contrast between the objects and the background. This is realized by positioning two light sources and a diffusing plate behind the container.

In order to obtain the trajectory of the sphere inside the sand, we attach a wire with markers which remain above the sand during impact and are imaged with the high-speed camera. This procedure is explained in greater detail in Section~\ref{boundaries}-A.

We use two ways to confine the impact and jet formation experiment. First of all, we study the influence of the bottom of the container by reducing the height at which the container is filled with sand down to a few ball diameters (Section~\ref{shallow_bed}). Second, to investigate the influence of the closeness of the side walls, we insert PVC cylinders of varying diameters into the sand, such that the cylinder axis coincides with the trajectory of the ball inside the sand. In this procedure sufficient care was taken that the presence of the cylinder did not disturb the fluidization and settling process of the sand bed (Section~\ref{boundaries}).

\begin{figure}[htp]
\includegraphics[width=\linewidth]{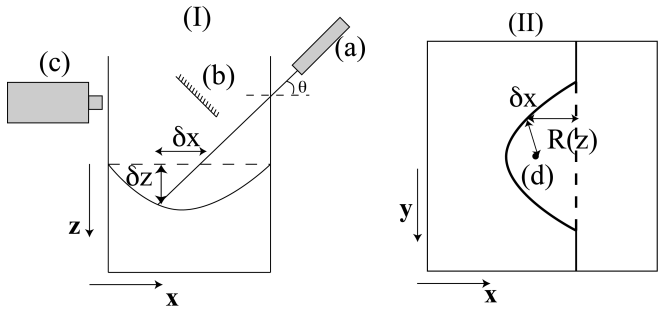}
\caption{(I) Laser profilometer. A diode laser sheet (a) is directed onto the surface at an angle $\theta$. Using a mirror (b) and a high-speed camera (c), images of the surface are recorded. (II) Schematic view of the resulting surface. The dashed line represents the laser sheet when the surface is flat and the continuous line the laser sheet when the surface is perturbed. The local deviation $\delta x = x(y) - x_l$ of the laser sheet is related to the vertical coordinate $\delta z = z(y)$ of the surface. (d) is the center of the cavity, from which the cavity radius $R(z)$ can be deduced.} \label{laser_setup}
\end{figure}

\begin{center}
\textbf{Time and position of collapse}
\end{center}
When traveling through the sand bed, the ball creates a cavity. The shape of the cavity is obtained using a profilometer similar to the one described in~\cite{refdeVet2007} (see Fig.~\ref{laser_setup}). A diode laser sheet with wavelength of $680$~nm strikes the granular media at an angle $\theta$, fixed arbitrarily at $55^{\circ}$. Using a mirror and a high-speed camera, we can measure the horizontal projection of the points where the laser sheet touches the sand from above. When the surface is flat, this projection is a straight line parallel to the $y$-direction; the coordinate of a point on this straight line is ($x_l$, $y$). When the surface is perturbed, the projection appears to be a curved line. For any point on this line with coordinate ($x(y)$, $y$)) the depth of the surface can be deducted --as a function of y-- from $x_l$ and $x(y)$
\begin{equation}
z(y) = (x(y)-x_l)\tan(\theta)\,.
\end{equation}
If we assume rotational symmetry of the cavity around the center of the ball [denoted as ($x_c$,$y_c$)] we can in addition deduce the radius of the cavity at all these depths $z(y)$
\begin{equation}
R(z(y))=\sqrt{(x(y)-x_c)^2+(y-y_c)^2}\,.
\end{equation}
By analyzing each of the high speed imaging recordings in this way, we can obtain the cavity profile $R(z,t)$ as a function of both depth $z$ and time $t$ (up to a certain maximum depth that is set by the laser sheet angle $\theta$).

\section{Influence of the bottom: A shallow bed\label{shallow_bed}}

Now that we have introduced the experimental setup, we will continue with the discussion of our results: In this Section we start with what is observed in a less-filled container (i.e., a shallow sand bed) and in the next Section proceed with the discussion of what happens when the diameter of the system is decreased.

Before turning to the case in which the proximity of the container bottom becomes important, let us first recall in table~\ref{results_unconfined} the results obtained in the usual unconfined case, here at $\textrm{Fr} = 70$ and ambient pressure: the container is large enough ($D = 14 $~cm $\gg d$) to avoid any effect of the surrounding walls and deep enough (the height of the sand bed, $h_{\rm bed}$ is around $30$~cm, that is $18.75\,d$) such that the bottom has no influence.

\begin{table}[h]
\begin{center}
\begin{tabular}{|c|c|c|}
\hline
\hline
Final depth $z_{\rm f}$ & Stopping time $t_{\rm stop}$ & collapse time $t_c$ \\
\hline
$11\,d$ & $110$~ms & $51$~ms\\
\hline
\hline
closure depth $z_c$ & jet height $h_{\rm jet}$ & eruption time $t_{\rm erup}$ \\
\hline
$2\,d$ & $18.5\,d$ & $510$~ms\\
\hline
\hline
\end{tabular}
\end{center}
\caption{Results obtained at $\textrm{Fr} = 70$ and $p = 1$~bar in the usual unconfined case, i.e., in a deep bed with $h_{\rm bed}=18.8\,d$. These values will be used as reference values in the discussion of the experimental results.}
\label{results_unconfined}
\end{table}

We modified the height of the sand bed, $h_{\rm bed}$ by simply adjusting the amount of sand in the container. The first and most conspicuous effect is that below a certain depth of the sand bed the impacting sphere is stopped abruptly by its collision with the container bottom, rather than slowly being stopped by drag as happens in the unconfined case. In this way, decreasing the depth of the sand bed allows us to look at the influence of the final depth of the ball, $z_{\rm f}$, and the cavity size on the jet and the eruption.

\subsection{Influence on the jet}

\begin{figure}[htp]
\centering
\includegraphics[width=\linewidth]{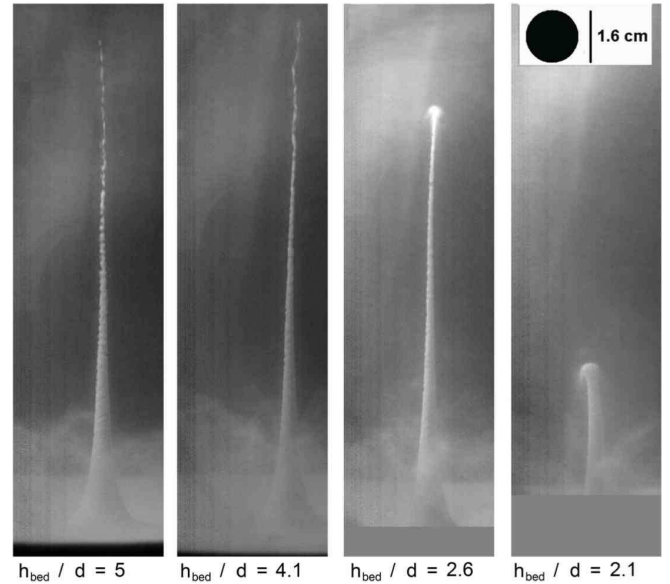}
\caption{Influence of the height of the sand bed $h_{\rm bed}$ on the shape and height of the jet for $\textrm{Fr} = 70$ and $p = 1$~bar: Images of the jet, taken at $0.12$~s after the ball impact for four different bed heights, decreasing from left to right. Below a threshold there is a clear change in height and width of the jet.}
\label{jets1000}
\end{figure}

\begin{figure}[htp]
\centering
\includegraphics[width=\linewidth]{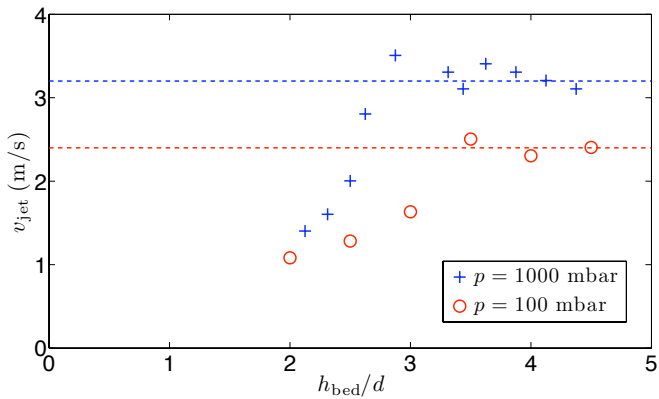}
\caption{Initial velocity of the jet, $v_{\rm jet}$ as a function of the height of the sand bed, $h_{\rm bed}$ for $\textrm{Fr} \sim 70$ and $p = 1000$~mbar ($+$) and $p = 100$~mbar ($\circ$). There is a sharp threshold below which the initial jet velocity rapidly decreases. The dashed lines represent the undisturbed values of $v_{\rm jet}$, measured in a deep bed ($h_{\rm bed}=18.8\,d$).}
\label{v_jetvs_h_bed}
\end{figure}

In Fig.~\ref{jets1000}, we show four images from the jet formed when the ball is dropped into the sand bed for $\textrm{Fr} = 70$ and ambient pressure. While reducing $h_{\rm bed}$, there is no change in the jet shape or height down to a certain threshold. Below this threshold, the jet reaches less high and becomes broader, most notably at the top. The maximum height of the jet, $h_{\rm jet}$, is obtained by measuring the initial jet velocity $v_{\rm jet}$ as soon as it appears above the surface of the sand (using energetic arguments, $h_{\rm jet} \propto v_{\rm jet}^2$). The initial  jet velocity $v_{\rm jet}$ is plotted as a function of $h_{\rm bed}$ in Fig.~\ref{v_jetvs_h_bed}: For $h_{\rm bed}$ higher than $3\,d$, $v_{\rm jet}$ saturates to its undisturbed value of approximately $3.2$ m/s. Reducing $h_{\rm bed}$ below $3\,d$, $v_{\rm jet}$ decreases rapidly. When we reduce the ambient pressure to $p = 100$~mbar, we find the same behavior (see Fig.~\ref{v_jetvs_h_bed}) although the crossover takes place at a slightly higher value of $h_{\rm bed}$. Remarkably, in both cases this decrease does not happen at the depth at which the ball is stopped by the bottom (which would be around $h_{\rm bed}\,=\,11\,d$ and $h_{\rm bed}\,=\,6\,d$ for $p\,=\,1000$~mbar and $p\,=\,100$~mbar respectively) but at a much lower depth of $h_{\rm bed}\,\approx\,3\,d$.

This can be explained as follows: The closure depth, $z_{c}$, remains unaltered by the presence of the bottom (which below $h_{\rm bed} = 11\,d$ only makes the ball stop earlier and less deep) until the bed depth becomes less than the sum between the position of the unconfined collapse ($2\,d$, see table~\ref{results_unconfined}) and the diameter of the ball. Below this value, the collapse happens on top of the ball leading to a less directional top of the jet which has a more or less spherical shape; moreover the closure depth decreases when the bed becomes smaller and so does the initial jet velocity.

\subsection{Influence on the eruption\label{shallow_eruption}}

\begin{figure}[htp]
\includegraphics[width=\linewidth]{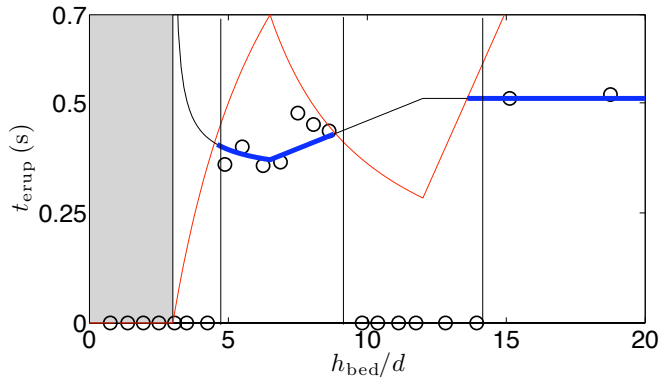}
\caption{The time $t_{\rm erup}$ when the granular eruption at the surface starts is plotted as a function of the height of the sand bed $h_{\rm bed}$, for $\textrm{Fr} = 70$ and $p = 1$~bar (black open circles). Measurement points with $t_{\rm erup} = 0$ correspond to those cases where no eruption was observed. The experimental regimes with and without eruption are separated by the vertical black lines. The grey region represents the region where no air bubble is entrapped. The thin blue and red lines represent the different time scales that are involved in the problem $t_1$ is the time the air bubble needs to reach the surface (black thin line) and $t_2$ is the time the air bubble needs to diffuse within the sand bed (red thin line). When $t_1$ is smaller than $t_2$, an eruption is expected; this is depicted by the continuous thick blue line. The different regions obtained from the timescale argument qualitatively correspond to the experimental results. More details about the way in which $t_1$ and $t_2$ are estimated are provided in the main text and in footnote~\cite{footnote2}.} \label{eruption_h_bed}
\end{figure}

Providing that the void collapse does not happen directly at the ball, an air bubble is entrapped. The volume of this bubble can be estimated as:
\begin{equation}
V_{\rm bubble} \propto h_{\rm rad}^2 (z(t_c)- z_c)\,\sim d^2 (z(t_c)- z_c)\,,
\label{eq_bubble}
\end{equation}
where $z(t_c)$ is the position of the ball at closure and $h_{\rm rad}$ is the radial length scale of the bubble, which can be approximated by the diameter of the ball. The bubble slowly rises through the sand and can lead to a violent granular eruption. However, this eruption is not always observed. To study when and why this is the case, in Fig.~\ref{eruption_h_bed} we plot the time between impact and eruption, $t_{\rm erup}$, as function of the height of the sand bed, $h_{\rm bed}$. (Note that measurement points with $t_{\rm erup} = 0$ correspond to those cases where no eruption was observed.)
Up to a certain threshold, which is around $4.8\,d$, no eruptions can be observed. This can be attributed to the fact that, while rising, small air bubbles just dissolve into the sand bed before reaching the surface. When the bed gets deeper, the air bubble reaches a certain critical volume $V^*$, above which a granular eruption can be seen. From the experimental results, this size found to be around $V^*\sim d^2(z(t_c)-z_c)\,\sim\,3.8\,d^3$. Then remarkably, above $9\, d$ the eruptions disappear again and reappears only when $h_{\rm bed}>14\, d$.

This peculiar behavior can be understood, at least qualitatively, from the competition of the two time scales corresponding to the two mechanisms the air in the bubble has to escape from the bed:
\begin{itemize}
\item The bubble needs a time $t_1$ to reach the surface. First of all, for $h_{\rm bed} < 3\,d$, the collapse happens on top of the ball, and no air bubble is entrapped. Between $3\,d$ and $5.5\,d$, the position of the ball at closure, $z(t_c)$, increases and so does the volume of the air bubble; in this region, $t_1$ decreases. While increasing the sand depth even further, the volume of the air bubble remains constant, but the initial position of the bubble is deeper since the entrapped bubble will follow the ball until it stops. Therefore $t_1$ will increase until $h_{\rm bed}$ is equal to $11 d$ which is the final depth of the ball in the unconfined case. Above this value, there is no change on the final depth and $t_1$ becomes independent of the depth of the sand bed. This is depicted by the thin black line in Fig.~\ref{eruption_h_bed}. More details about the way that $t_1$ is estimated are given in~\cite{footnote2}.
\item The air in the bubble trapped by the collapse escapes --in the dissolution time $t_2$-- through pressure driven flow through the porous bed. Factors that affect this process are the size of the bubble (which determines the amount of air that needs to escape), the pressure of the air (which approximately equals the hydrostatic pressure in the surrounding sand), and the length of the path the air needs to travel. For this last quantity we need to consider that air can both escape through the top and through the bottom of the bed, the latter due to the presence of the sintered plate. These factors combine into Darcy's law: $Q \propto \Delta P/H$, where the flow rate $Q$ depends on the pressure difference $\Delta P$ driving the flow and the path length $H$. Turning to Fig.~\ref{eruption_h_bed} again, for $h_{\rm bed} < 3\,d$, no air bubble is entrapped. Between $3\,d$ and $5.5\,d$, $z(t_c)$ --and therefore the bubble size-- increases, leading to a steep increase of the dissolution time $t_2$. Upon increasing the sand depth beyond $5.5\,d$ the bubble size remains constant but the pressure inside the bubble increases. From Darcy's law we thus find that $t_2$ decreases. This continues until $h_{\rm bed}$ is equal to $11\,d$ beyond which the ball does not reach the bottom of the plate anymore. Note that until this point the shortest (and therefore chosen) path for the bubble to dissolve is towards the bottom of the container. If we now increase  $h_{\rm bed}$ beyond $ 11\,d$ this shortest path starts to grow, and with the path, using Darcy's law, also the dissolution time. This is captured by the thin red line in Fig.~\ref{eruption_h_bed}. More details about the estimation of $t_2$ are provided in~\cite{footnote2}.

\end{itemize}
As a result, an eruption can be only seen if the time $t_2$ becomes larger than $t_1$.  This is expressed by the continuous thick blue line, in qualitative agreement with the experimental behavior.

\section{Influence of the side walls\label{boundaries}}

\begin{figure*}
\centering
\includegraphics[width=0.85\linewidth]{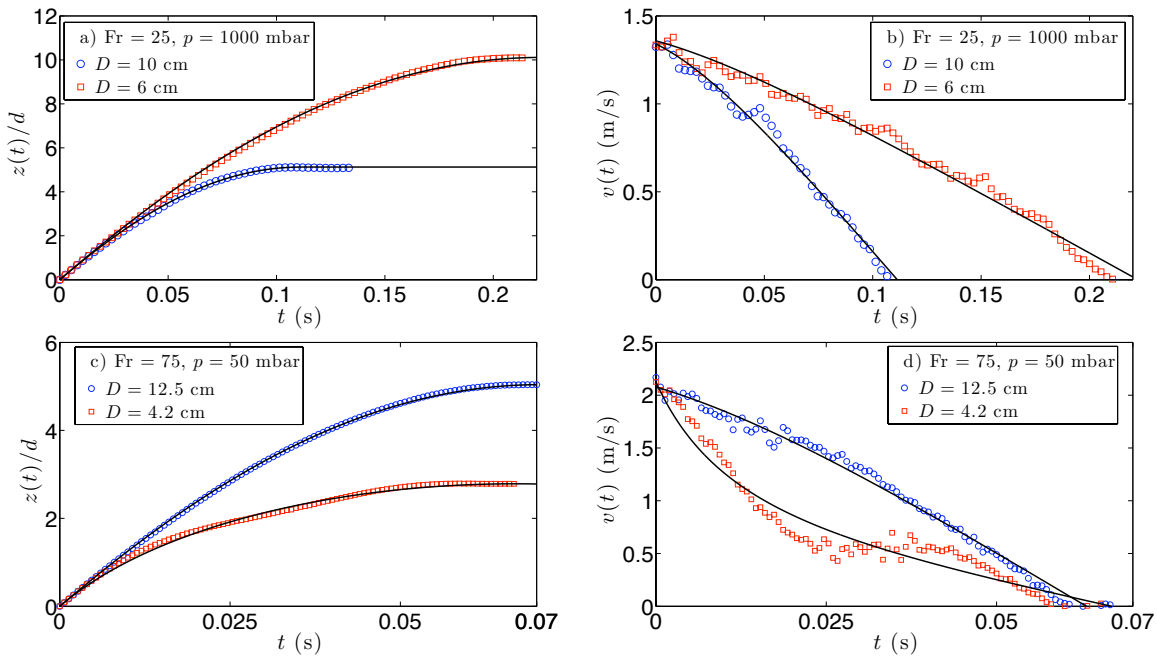}
\caption{\emph{(top)} Depth of the ball $z(t)$ (a) and its velocity $v(t)$ (b) as a function of time after impact for $\textrm{Fr} = 25$, $p = 1000$~mbar, $D = 10$~cm ($\circ$) and $D = 6$~cm ($\square$). The lines correspond to a fit using equation~\ref{eq:balistic} with $\kappa = 4.525$~N/m and $\alpha = 0.132$~kg/m for $D = 10$~cm and  $\kappa = 1.695$~N/m and $\alpha = 0.118$~kg/m for $D=6$~cm. \emph{(bottom)} Depth of the ball $z(t)$ (c) and its velocity $v(t)$ (d) as a function of time after impact $t$ for $\textrm{Fr} = 75$, $p=50$~mbar $D = 4.2$~cm ($\circ$) and $D = 12.5$~cm ($\Box$). Again, the lines correspond to a fit using equation~\ref{eq:balistic} with $\kappa = 14$~N/m and $\alpha = 0.281$~kg/m for $D = 4.2$~cm and with  $\kappa = 13.5$~N/m and $\alpha = 0.111$~kg/m for $D=12.5$~cm. Within the smallest container, and only at low pressure, we observe anomalous behavior: The ball reaches a plateau in which the velocity remains constant before going to zero again at larger times. Clearly, the model fails to describe the data in this case.}
 \label{fig:trajectory}
\end{figure*}

In the previous section, we discussed the influence of the bottom of the cavity on the process of object penetration and jet formation and found that, if the sand depth is fixed at $30$~cm, there is no effect of the bottom on the jet formation process. Fixing this bed depth, we now turn to study the effects of the side walls of the container on the complete series of events leading to the jet. For this, some cylinders of different diameters $D$ are placed inside the sand during the fluidization process: we choose $D = 4.2$~cm, $6$~cm, $8.5$~cm, $10$~cm and $12.5$~cm. In this way we change the aspect ratio $D/d$ from $2.6$ to $7.8$.

\subsection{Ball trajectory}

The first thing that happens upon impact of the ball onto the surface is that it penetrates and creates a void inside the sand bed. The question we ask in the next subsection concerns the influence of the container diameter on the drag force experienced by the ball during its motion through the bed. To monitor the trajectory of the ball, a wire with a tracer is attached to the ball. Using a high-speed camera ($1500$ frames per second) and image analysis, we obtain the trajectory of the tracer and therefore the trajectory of the ball $z(t)$. To keep the wire tense an extra friction device and a light counterweight are used , which have the effect that the ball experiences a downward acceleration due to gravity which is approximately 10 \% smaller than $g$.  The actual acceleration is measured during the ``free fall'' part of the trajectory, and the results presented here have been corrected for this effect.

In the top two plots of Fig.~\ref{fig:trajectory}, we compare the trajectories of the ball at ambient pressure for an impact with $\textrm{Fr} = 25$ and for two diameters of the confining cylinder ($D = 6.0$ and $10.0$ cm). We can fit the experimental trajectories using the model introduced in Section~\ref{Theory} (Eq.~\ref{eq:balistic}) using $\alpha$ and $\kappa$ as fitting parameters. The agreement between the model and the experiments is very good (see Fig.~\ref{fig:trajectory}).

\begin{figure}[htp]
\includegraphics[width=0.9\linewidth]{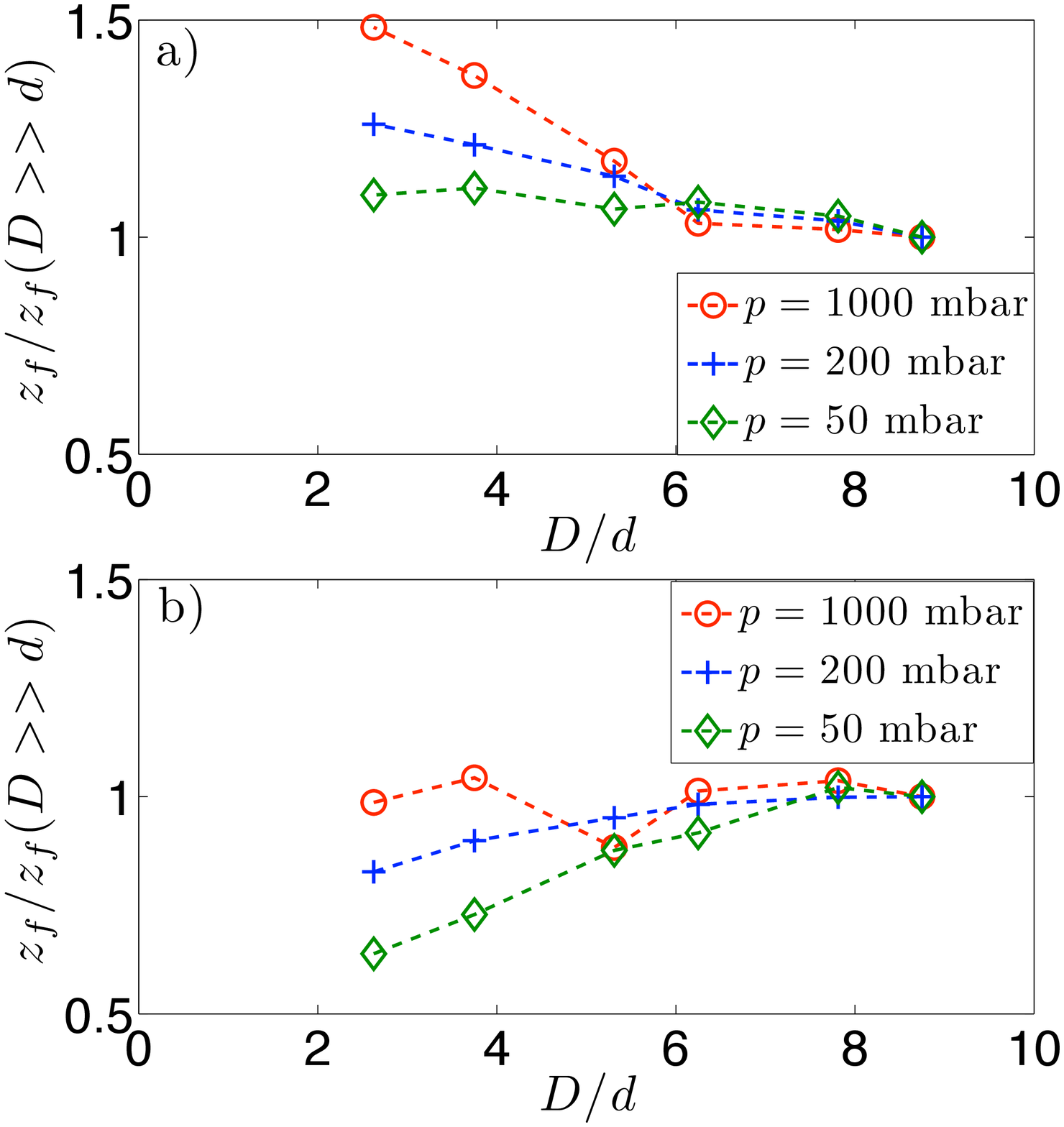}
\caption{Final depth $z_{\rm f}$ as a function of the container diameter $D$, at different pressures,for a) $\textrm{Fr} = 25$ and b) $\textrm{Fr} = 75$. The final depth is divided by the final depth for the unconfined case in order to emphasize the deviations due to the proximity of the boundaries. The dashed lines are a guide to the eye to separate the different pressures.}
 \label{finaldepths}
\end{figure}

Decreasing the diameter of the container surprisingly increases both the final depth of the ball, $z_{\rm f}$ and the time to reach the final depth, $t_{\rm s}$. In Fig.~\ref{finaldepths}a, we report the final depth $z_{\rm f}$ as a function of the container diameter at different pressures for $\textrm{Fr}=25$. There is a clear dependance: The final position of the ball is deeper for a smaller container. Also, the influence of the boundaries for this Froude number is less pronounced at small pressures. We conclude that for $\textrm{Fr}=25$ the drag force the ball experiences becomes smaller for small containers.

But what happens at higher Froude numbers? In Fig.~\ref{finaldepths}b, we report the final depth, $z_{\rm f}$ as a function of the container diameter for $\textrm{Fr} = 75$. At first glance the behavior now seems completely opposite to what we observe at small Froude number, as the final depth now decreases with decreasing container diameter: To be more precise, at atmospheric pressure the final depth stays more or less constant and at lower pressures there is a decrease in $z_{\rm f}$ with decreasing container diameter. So now the drag force seems to be larger for small container diameters.

\begin{figure}[htp]
\includegraphics[width=0.9\linewidth]{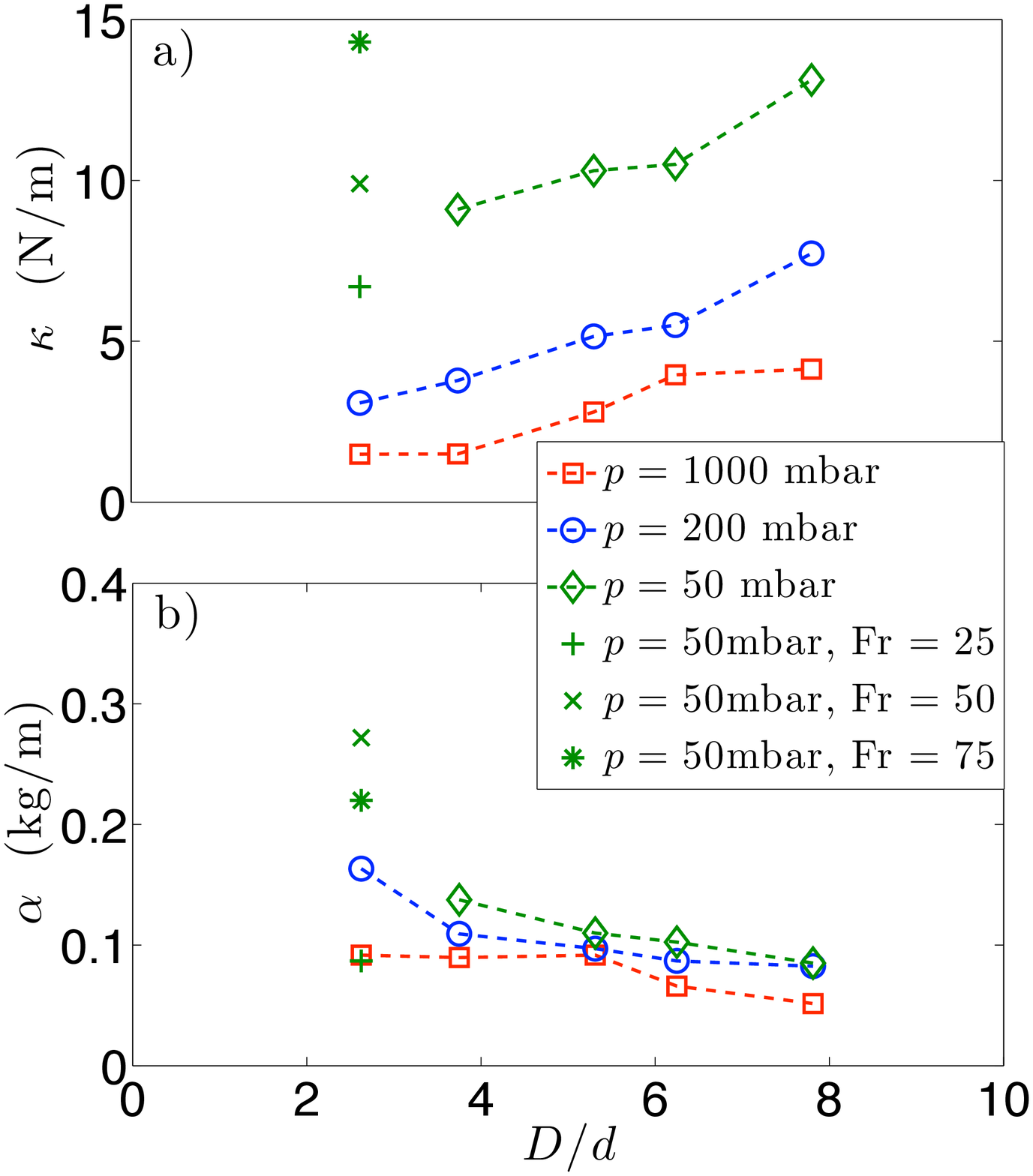}
\caption{ a) $\kappa$ and b) $\alpha$ as a function of cylinder diameter $D$ for different pressures $p$. For almost all values of $D$ and $p$ variations of both $\kappa$ and $\alpha$ are within the measurement error and each point is obtained from an average over a range of Froude numbers from $25$ to $100$. Only for the smallest container ($D/d=2.6$) and the lowest pressure ($50$ mbar), there is a strong dependance of $\kappa$ and $\alpha$ on the Froude number; the model is not valid in this situation. Plot b) reveals that for large Fr the quadratic drag takes over for small cylinder diameters leading to less intrusion of the ball (Fig.~\ref{finaldepths}b).}
 \label{kappa_FR_25}
\end{figure}

To understand this difference, we have to separately look at the behavior of the hydrostatic and quadratic 
 drag force:  After all, for small Froude numbers we expect that the hydrostatic drag $-\kappa z$ will dominate the behavior of the ball, whereas for higher impact velocities it is expected that the quadratic drag $\alpha v^2$ will start to become increasingly more important during the motion of the ball. To this end, in Fig.~\ref{kappa_FR_25} we plot $\kappa$ and $\alpha$ as a function of container diameter for three different pressures. Each value represents the average parameters obtained from fits to the trajectory data analogous to the ones of Fig.~\ref{fig:trajectory} over a range of Froude numbers from $25$ to $100$ \cite{footnote3}. As shown by Caballero~\cite{refCaballeroPRL07}, the hydrostatic force depends on the ambient air pressure: $\kappa$ decreases with $p$ roughly as $p^{-1/2}$. Our findings are consistent with this observation, also for smaller container diameters (not shown). Next to this we find that $\kappa$ increases quite steeply with $D$, which is consistent with the lower drag experienced by the impacting ball at smaller container diameters at low Froude numbers. Physically, the behavior of the  hydrostatic drag force can be understood using a similar argument as~\cite{refCaballeroPRL07}: When the ball passes through the sand, an air flow is created around it which fluidizes the sand bed and reduces the drag force. This effect is expected not only to be more important at higher pressure but also when the container diameter becomes smaller: Near the wall, the velocity of the interstitial air is required to be zero and, since the same amount of air needs to be displaced, the flow will be more important if the aspect ratio $D/d$ is small. Consequently, the  hydrostatic drag force will be lower.

Figure~\ref{kappa_FR_25}b contains the coefficient $\alpha$ of the quadratic drag term $\alpha v^2$. Clearly, $\alpha$ becomes larger for smaller container diameters but the difference is hardly as pronounced as was the case for $\kappa$. This accounts for the observation that at some point, for larger Froude number, the drag does become larger when the container diameter is decreased.

Finally, in Figs.~\ref{kappa_FR_25}a and b there is one exceptional value: For the smallest container diameter ($D/d=2.6$) and the lowest pressure ($50$ mbar) the fitted values of $\kappa$ and $\alpha$ turn out to strongly depend on the Froude number. The bottom two plots in Fig.~\ref{fig:trajectory}, which contain two trajectories at $50$~mbar for the largest and the smallest container diameter, reveal the reason why: Whereas for the biggest container ($D =12.5$~cm), the behavior is similar to the behavior described for $\textrm{Fr} = 25$, for the smallest one ($D = 4.2$~cm) it is qualitatively different. Whereas the agreement between the experimental and the computed trajectory sill seems to be reasonable (Fig.~\ref{fig:trajectory}c), the velocity of the ball (Fig.~\ref{fig:trajectory}d) presents large discrepancies: The measured ball velocity doesn't decrease to zero gradually, but first slows down until it reaches a plateau at constant velocity where it stays for a while before slowing down until it stops. This behavior is identical to the one observed in the X-ray experiments of Royer~\emph{et~al.}~\cite{refRoyerPRE08}, in which the container needed to be kept small. That behavior therefore is likely to be a boundary effect. We believe that the origin of the plateau lies in a depth-independent force between the ball and the wall (mediated by force chains) which is dominant over the hydrostatic drag force and, together with the quadratic drag force, balances gravity at the plateau velocity. At some depth, the Coulomb drag force takes over, slowing the ball down to zero. Obviously, the model cannot be valid in this situation and an extra force due to the ball/wall interaction should be taken into account.

\subsection{Collapse of the cavity}

\begin{figure}[htp]
\includegraphics[width=\linewidth]{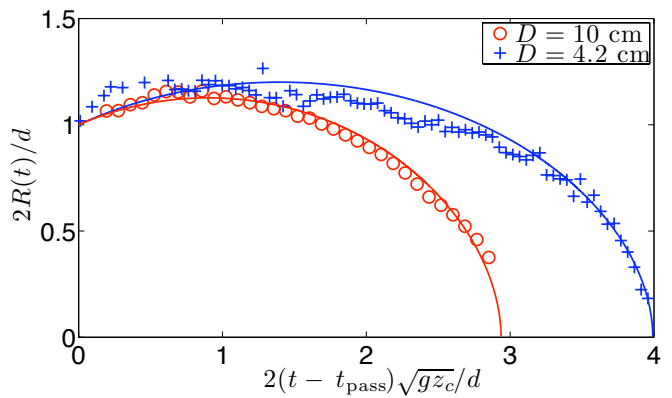}
\caption{Dynamics of the cavity collapse at closure depth for two container diameters $D = 4.2$~cm ($\Box$) and $D=10$~cm ($\circ$). Here, $\textrm{Fr} = 70$ and $p = 1$~bar. The time has been rescaled by multiplying with a factor $2\sqrt{gz_c}/d$ in order to show the results in a single plot. The continuous line correspond to a fit using the 2D Rayleigh-Plesset equation (Eq.~\ref{eq:Rayleigh}).}
\label{cavity_dynamic}
\end{figure}

The second issue that we want to address in this Section is the influence of the container diameter on the collapse of the cavity. We study the dynamics of the collapse of the cavity at closure depth using the profilometric method described in detail in Section~\ref{exp_setup}. In Fig.~\ref{cavity_dynamic}, the radius of the cavity is plotted as a function of time $t-t_{\rm pass}$ for two different diameters at atmospheric pressure where $t_{\rm pass}$ is the time needed for the ball to reach the closure depth $z_c$. We can clearly distinguish a slight expansion of the cavity followed by a 
strong contraction. The collapse accelerates towards the pinch-off. Due to experimental limitations we do not have enough spatial resolution to obtain data points close to the pinch-off. The void dynamics is in qualitative agreement with the behavior predicted by the 2D Rayleigh-Plesset equation described in section~\ref{Theory} (Eq.~\ref{eq:Rayleigh}). Whereas the expansion turns out to be weak and more or less independent of the container diameter, the contraction and the closure strongly depend on it. A plausible explanation would be that, for small containers, less sand is involved in the collapse. Therefore, the hydrostatic pressure which drives the collapse is not as sustained as for a larger container, explaining why the collapse takes longer for a smaller container (Fig.~\ref{cavity_dynamic}).

\begin{figure}[htp]
\includegraphics[width=\linewidth]{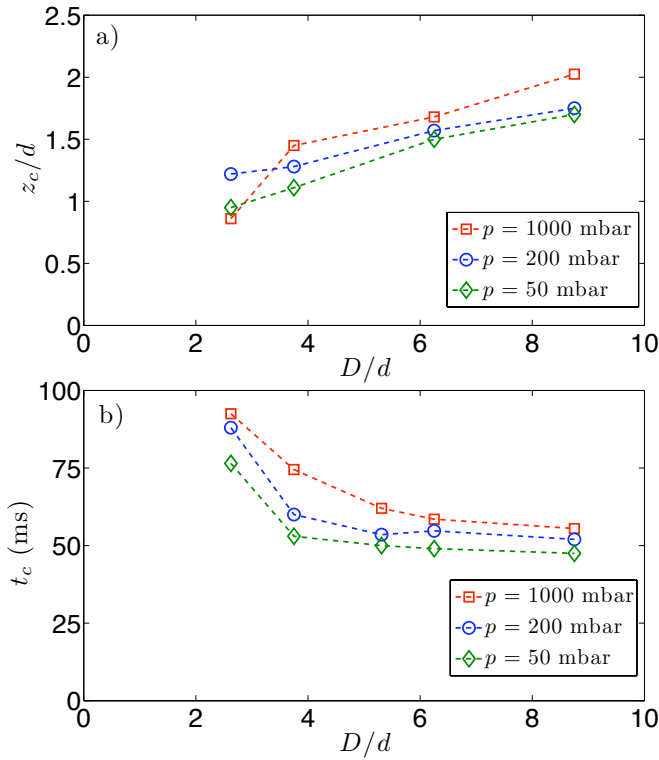}
\caption{(a) Closure depth $z_c$ as a function of the container diameter $D$ for different pressures. (b) Closure time $t_c$ as a function of the container diameter $D$ for different pressures. For all measurements $\textrm{Fr} = 70$.}
\label{collapse_Fr=70}
\end{figure}

In Fig.~\ref{collapse_Fr=70} we plot the closure depth $z_c$ and the closure time $t_c$. We find that $t_c$ increases and $z_c$ decreases when decreasing the container diameter. This decrease of the closure depth is generic: Also for small $\textrm{Fr}$, where $z_{\rm f}$ actually increases, we find a decrease of $z_c$.
The fact that a decrease of the closure depth $z_c$ implies an increase of the collapse time $t_{\rm coll}$ can be understood from a reduction of the driving pressure ($\propto gz_c$) and the availability of less sand for smaller container diameters (as explained above).

\subsection{Jet Height}

\begin{figure}[htp]
\includegraphics[width=\linewidth]{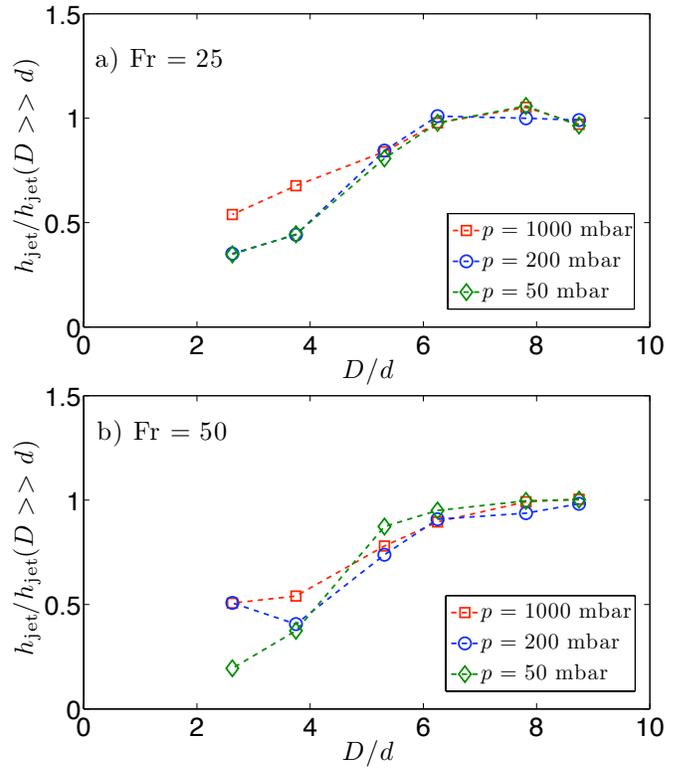}
\caption{The jet height, $h_{\rm jet}$ as a function of the container diameter $D$ for $\textrm{Fr} = 25$ (a) and $\textrm{Fr} = 50$ (b) at different ambient pressures. The jet height is divided by the jet height in the unconfined case in order to see the deviations due to the proximity of the boundaries. For all pressures and Froude numbers the jet height increases with increasing container diameter. The dashed lines are a guide to the eye to separate the measurement series at different pressures.}
\label{jetheights_diameter25}
\end{figure}

\begin{figure}[htp]
\includegraphics[width=\linewidth]{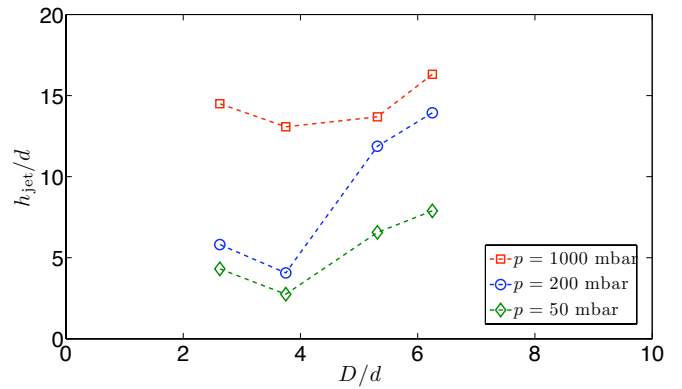}
\caption{The jet height $h_{\rm jet}$ as a function of the container diameter $D$ for Froude = 100 and different ambient pressures. Again, there is a clear change in jet height as function of container diameter. Measurements at the highest Froude numbers are not possible due to the surface seal (see text). The dashed lines are a guide to the eye to separate the measurement series at different pressures.}
\label{jetheights_diameter100}
\end{figure}

Now that we studied how the process of the formation and subsequent collapse of the cavity are influenced by the proximity of the side walls, we continue with the influence of the diameter of the container $D$ on the jet and, in particular, on the maximum height reached by the jet. In Fig.~\ref{jetheights_diameter25}, the jet height $h_{\rm jet}$ is plotted as a function of the diameter for two Froude numbers ($\textrm{Fr}=25$ and $\textrm{Fr}=50$) at different values of the ambient pressure.  Since it was already discussed extensively in~\cite{refCaballeroPRL07} that the jet becomes smaller at reduced ambient pressure, we now divide $h_{\rm jet}$ by the jet height in the unconfined situation. We observe that, while decreasing the container diameter, the jet reaches less high. This behavior is the combined result of the reduction of the closure depth and the increase of the closure time with decreasing container diameter as was described in the previous subsection: The reduction of $z_c$ reduces the hydrostatic pressure that drives the collapse and the increase of the closure time is connected to the fact that --because of the reduced container diameter-- there is less sand available during the collapse, making the driving pressure less sustained. Both factors contribute to a decrease of the jet height. The rescaling by the unconfined jet height also reveals that the influence of the boundaries is similar for all pressures and even for these two different Froude numbers. The unconfined behavior is obtained when the diameter of the container is seven times higher than the diameter of the ball.

At high Froude number (Fr = $100$), the results can only be obtained for small containers, because, when the diameter is large, the jet collides with the splash which is being sucked into the cavity behind the ball at high ambient pressures. This is similar to the surface seal that has been observed for impacts on a liquid \cite{refGilbarg48,refBergman09}. For this high Froude number the results are less 
conclusive, as can be seen in Fig.~\ref{jetheights_diameter100}. This is possibly due to the increased importance of the air flow caused by the ball when it is restricted to a smaller container diameter at these high impact velocities.

\subsection{Granular eruption}

\begin{figure}[htp]
\includegraphics[width=\linewidth]{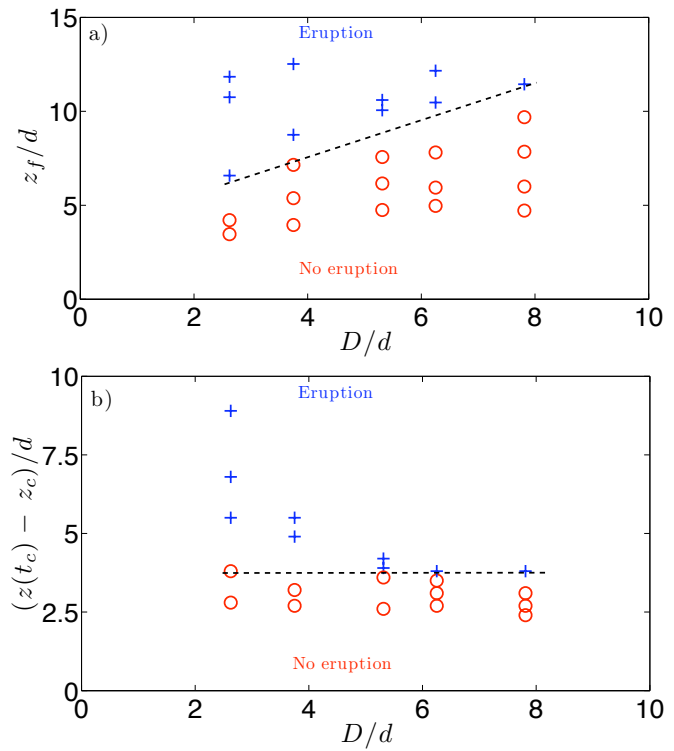}
\caption{(a) Phase diagram for the granular eruption at $\textrm{Fr} = 100$ as a function of the final depth $z_{\rm f}$ and the container diameter $D$. In both plots red open circles indicate parameter values where an eruption was absent, whereas blue plus signs stand for parameter values with an eruption. Note that $z_{\rm f}$ (which is a measured quantity) has been varied by using different values of the ambient pressure $p$. (b) The same phase diagram, now as a function of the volume of the entrapped air bubble ($(z(t_c)-z_{\rm c})/d$) and the container diameter $D$. The latter plot clearly indicates that the presence of the eruption is a function of the entrapped air bubble size only \cite{footnote4}.}
\label{eruptionfr100}
\end{figure}

Finally, we turn to the granular eruption that terminates the series of events. Since the container diameter has an influence on both the final depth and the closure depth, it is expected that the granular eruption will depend on the container diameter $D$. In Fig.~\ref{eruptionfr100}a we report, for $\textrm{Fr}=100$, the phase diagram indicating the presence of an eruption in $(z_{\rm f},D)$-space. Note that the different values for the final depth $z_{\rm f}$ at fixed container diameter $D$ have been obtained by varying the ambient pressure $p$. There is a marked dependance on the container diameter $D$: More eruptions are observed in a small container than in a large container.

This behavior can be explained using the influence of the side walls on the trajectory of the ball and on the collapse time together with the closure depth: For the same pressure, the closure time is larger, which leads to a deeper position of the ball at closure $z(t_c)$, and at the same time the closure depth is smaller, increasing the size of the entrapped air bubble for small container diameters. If we replace the final depth in Fig.~\ref{eruptionfr100}a by the quantity $(z(t_c)-z_c)/d$ which is proportional to the volume of the entrapped air bubble [remember that it was argued that $V_{\rm bubble} \sim d^2(z(t_c)-z_c)$, see Eq.~(\ref{eq_bubble}), which is subsequently divided by $d^3$], we remove the dependance on $D$:
In Fig.~\ref{eruptionfr100}b the phase diagram is separated into two parts using a horizontal line representing a critical volume $V^* \sim 3.8 d^3$. This means that, independently of the diameter of the container, the bubble volume upon its formation has to be big enough to lead to an eruption. As was explained in Section~\ref{shallow_bed} this is because the air bubble must have sufficient time to reach the surface before it has completely dissolved into the sand bed. Incidentally, the value for the critical volume determined from the phase diagram corresponds well to the value found in Section~\ref{shallow_bed}.

\section{Jet shape and thick-thin structure}
\label{TTshapes}

\begin{figure}[htp]
\includegraphics[width=\linewidth]{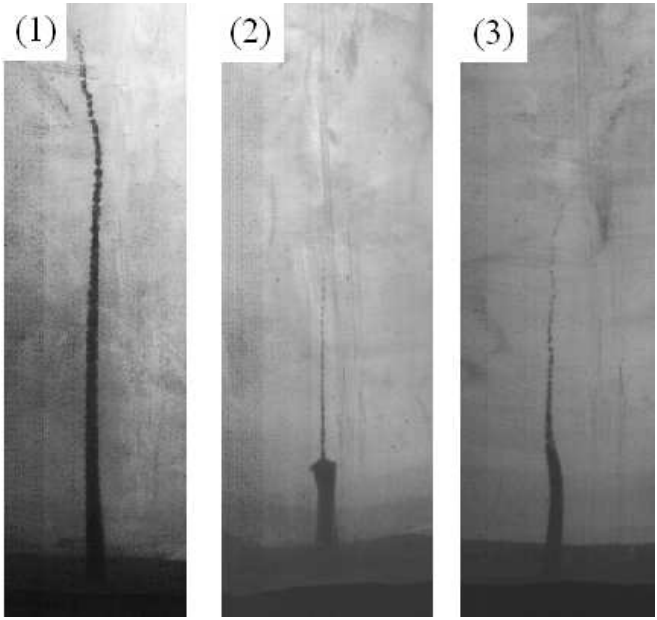}
\caption{Typical snapshots of the three distinct jet shapes observed in experiment: 1) Normal jet (for $D = 10$~cm, $\textrm{Fr} = 100$ and $p = 1000$~mbar); 2) Thick-thin structure with sharp shoulder (for $D = 8.5$~cm, $\textrm{Fr} = 100$ and $p = 100$~mbar); 3) Thick-thin structure with a transition (for $D = 10$~cm, $\textrm{Fr} = 50$ and $p = 50$~mbar). All snapshots show the fully developed shape of the jet at its maximum height. The snapshots are not on the same scale.}
\label{3jetshapes}
\end{figure}

The proximity of the side walls and the bottom does not only affect the height of the jet but also its shape. One of the most prominent features is the thick-thin structure first described by Royer {\it et al.}~\cite{refRoyerNP05,refRoyerPRE08} who studied the dependence of this structure on ambient pressure and Froude number. In the same work Royer {\it et al.} proposed a formation mechanism for the thick part of the jet based on the pressurized air bubble pushing sand into the thin jet originating from the pinch-off at the closure depth.

In this Section we report, in addition to the Froude and pressure dependence, a pronounced dependence of the thick-thin structure on the proximity of the container boundaries. We propose an alternative model for the formation of the structure which semi-quantitatively accounts for the observed behavior of the phenomenon for the entire parameter space.

In our experiments we can distinguish three different jet shapes, two of which exhibit a thick-thin structure:
\begin{enumerate}[(1)]
\item a 'normal' jet, in which the width of the jet gradually decreases from bottom to top,
\item a thick-thin structure with a sharp shoulder, where the thick lower part abruptly changes into a thin upper part,
\item a thick-thin structure with a transition, characterized by a transient region in which the thick lower part gradually passes into the thin upper part.
\end{enumerate}
An example of each of the three jet shapes is shown in Fig.~\ref{3jetshapes}.

\begin{figure}[htp]
\centering
\includegraphics[width=\linewidth]{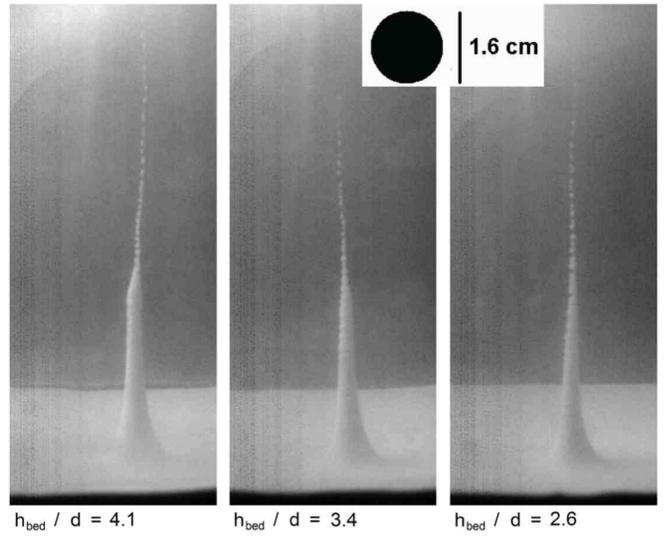}
\caption{Three snapshots of the shape of the jet at different values of the height $h_{\rm bed}$ of the sand bed, taken $120$~ms after the ball impact for $P = 100$~mbar and $\textrm{Fr} = 70$. For $h_{\rm bed} = 4.1 \, d$ there is a clear thick-thin structure (with a transition region), which gradually disappears when the bed height is decreased to $3.4 \, d$ and $2.6 \, d$.} \label{jets100}
\end{figure}

First, we briefly look at the influence of the bed depth on these structures for a moderate Froude number ($\textrm{Fr} = 70$). At atmospheric pressure we observe a `normal' jet for all values of the bed depth $h_{\rm bed}$ (Fig.~\ref{jets1000}). To observe a thick-thin structure we need to go to smaller ambient pressures: At $100$~mbar, a thick-thin structure with sharp shoulders can be observed in the unconfined case, i.e., for large $h_{\rm bed}$ (Fig.~\ref{3jetshapes}). Below a certain threshold ($h_{\rm bed} \leq 4 \, d$), the thick-thin structure gradually disappears (Fig.~\ref{jets100}). This disappearance coincides with the disappearance of the entrapped air bubble below $3 \, d$ in which case the collapse happens more or less on top of the ball. 

\begin{figure}[htp]
\includegraphics[width=0.9\linewidth]{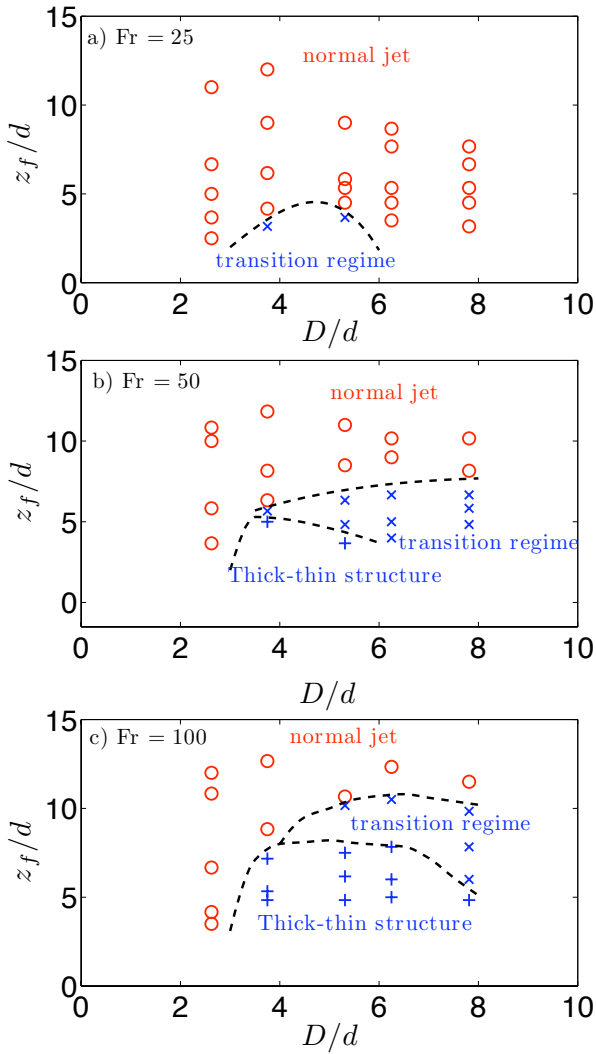}
\caption{Phase diagram of the observed jet shapes as a function of container diameter $D$ and final depth $z_{\rm f}$ for three different Froude numbers:  (a) $\textrm{Fr}=25$; (b) $\textrm{Fr}=50$; and (c) $\textrm{Fr}=100$. The different final depths  $z_{\rm f}$ (at fixed $D$) correspond to different ambient pressures ($ p = 50$,  $100$, $200$, $500$, and $1000$~mbar). The dashed lines are a guide to the eye to separate the different regions in the phase diagrams.}
\label{thickthin}
\end{figure}

The effect of the proximity of the side walls (within a sufficiently deep bed) is reported in the three phase diagrams of Fig.~\ref{thickthin}, where the jet shapes are classified as a function of container diameter and final depth \cite{footnote5}, for three different Froude numbers. For the lowest Froude number ($\textrm{Fr} = 25$), a thick-thin structure with a transition is found only for the smallest $z_{\rm f}$ (which corresponds to the lowest pressure, $p = 50$~mbar) at intermediate container diameter. Thick-thin structures with a sharp shoulder are not found for this Froude number. When we increase the Froude number, the thick-thin-structure region is found to grow. Within the region containing the transition variety of the thick-thin structure we observe the formation and growth of a region containing the sharp-shoulder variety. Although the thick-thin-structure region grows to include the largest container diameters that we have used in our experiment \cite{footnote6}, thick-thin structures are never found in the smallest container diameter for the parameter space explored in this study.

Now, which mechanism causes these structures? To answer this question we hypothesize a second collapse that happens on top of the ball forming a second jet. Such a second collapse can be motivated from experiments in a quasi-twodimensional setup~\cite{refMikkelsen02} and from X-ray measurements~\cite{refRoyerNP05,refRoyerPRE08}, where multiple collapses have also been observed. The idea is as follows: Since the second collapse happens at a later point in time, the first jet is already well on its way in the formation process when the second one is being formed. Now, if the second jet can catch up with the first fast enough, it will hit its base and produce a thick-thin structure. When the time span between the two jets is too long however, the first jet will have (almost) fully formed and the collision of the second jet with its base will not disturb its shape \cite{footnote7}.

\begin{figure}[htp]
\centering
\includegraphics[width=\linewidth]{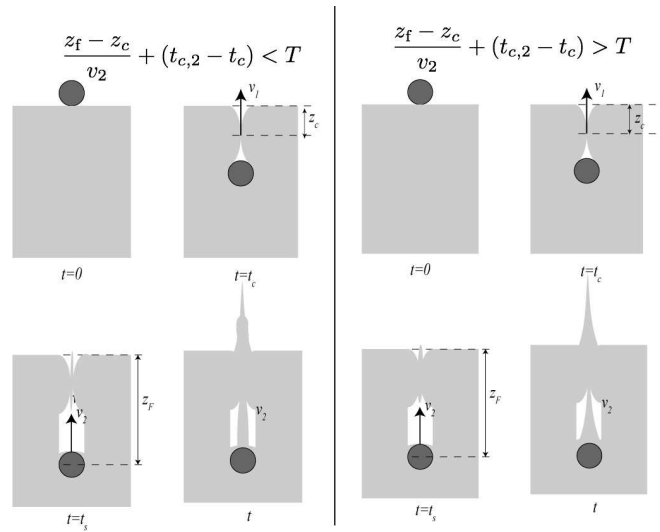}
\caption{Schematic drawing of the mechanism leading to the thick-thin structure. In case (a), the second collapse happens before a certain threshold time, such that the thickness of the layer of sand from the first collapse still is thin enough to be pushed up by the second jet and a thick-thin structure emerges. In case (b) we are above the threshold: The second jet collides with a thick layer of sand and is unable to disturb the formation of the first jet.} \label{schema}
\end{figure}

To quantify this idea we need to estimate the interval between the time that the first jet is formed at the closure depth $z_c$ and the moment that the second jet reaches $z_c$. This interval consists of the difference between the two closure times $(t_{c,2} - t_c)$ (where $t_{c,2}$ is the closure time of the lower collapse), summed with the time the second jet needs to reach $z_c$, i.e., $(z_{\rm f}-z_c)/v_2$ with $v_2$ the velocity of the second jet. If this time interval is shorter than some threshold value $T$, we obtain a thick-thin structure, as visualized in Fig.~\ref{schema}. This leads to:
\begin{equation}
\frac{z_{\rm f} - z_c}{v_2} + (t_{c,2} - t_c) < T
\label{timescaleargument}
\end{equation}

Let us illustrate the workings of this mechanism in an example: For $\textrm{Fr}=75$ and $p = 50$~mbar we start from the largest container size where a thick-thin structure is visible. When decreasing the size of the container, the closure depth $z_c$ and the final depth $z_{\rm f}$ decrease following approximately the same behavior, such that the distance between the two collapses is more or less constant. Because $_z{\rm f}$ decreases, the hydrostatic pressure and therefore the velocity of the second jet decrease as well, such that the first term in Eq.~(\ref{timescaleargument}) increases. The same holds for the second term, because the closure time is found to increase with decreasing container diameter (cf. Fig~\ref{collapse_Fr=70}b). Thus, the left hand side of Eq.~(\ref{timescaleargument}) increase with decreasing the container diameter, explaining why below a certain diameter the thick-thin structure disappears.

To check wether the argument of Eq.~(\ref{timescaleargument}) also works quantitatively we approximate the several terms in the equation with experimentally known quantities. In the first term $v_2$ is proportional to the square root of the driving hydrostatic pressure at depth $z_{\rm f}$, i.e., $v_2 = C \sqrt{g z_{\rm f}}$, with $C$ constant. Because, similarly, for the velocity of the first jet we have $v_{\rm jet} = C \sqrt{g z_c}$, we find $v_2 \approx \sqrt{z_{\rm f}/z_c} \, v_{\rm jet}$ which is inserted into the first term of Eq.~(\ref{timescaleargument}). In turn, $v_{\rm jet}$ can be deduced from the jet height $h_{\rm jet}$ as $v_{\rm jet} = \sqrt{2gh_{\rm jet}}$.

In the second term, the unknown quantity is the second closure time $t_{c,2}$ --i.e., of the cavity just above the ball-- which consists of the sum of the time $t_{\rm s}$ the ball needs to come to a standstill and the time $t_{\rm coll,2}$ the cavity needs to collapse at that point. Since according to the Rayleigh model discussed in Section~\ref{Theory} the collapse times should scale as $t_{\rm coll,2} = C' d/(2\sqrt{g z_{\rm f}})$ and $t_{\rm coll} = C' d/(2\sqrt{g z_c})$ respectively (with $C'$ constant), we have $t_{\rm coll,2} \approx \sqrt{z_c/z_{\rm f}} \, t_{\rm coll}$. Inserting all of the above in Eq.~(\ref{timescaleargument}) we obtain
\begin{equation}
\left[ \frac{z_{\rm f} - z_c}{\sqrt{2gh_{\rm jet}}} + t_{\rm coll} \right] \sqrt{\frac{z_c}{z_{\rm f}}} + (t_{\rm s} - t_c) < T\,.
\label{timescaleargument2}
\end{equation}

\begin{figure}[htp]
\centering
\includegraphics[width=\linewidth]{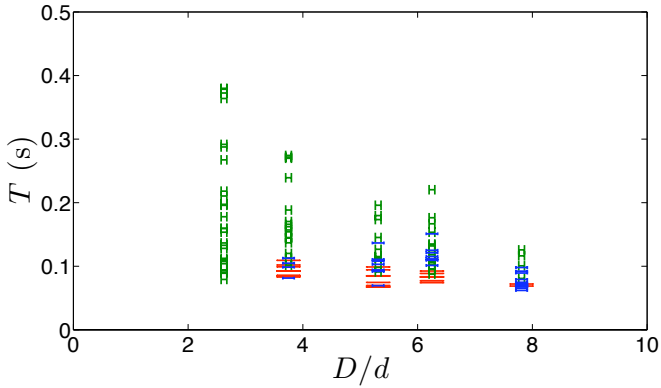}
\caption{Phase diagram with on the vertical axis the left hand side of Eq.~(\ref{timescaleargument2}) and on the horizontal axis the container diameter $D$. The plot contains all measurements from Fig.~\ref{thickthin}. Short, black dashes indicate normal jets, intermediate, blue dashes the thick-thin structure with a transition, and long, red dashes thick-thin structures with a shoulder.} \label{fig19}
\end{figure}

In Fig.~\ref{fig19} we find a phase diagram in which all measurements from Fig.~\ref{thickthin} are plotted again, but now with the left hand side of Eq.~(\ref{timescaleargument2}) on the vertical axis. Clearly, all thick-thin structures (intermediate and large dashes) lie below some time-threshold, in agreement with the formation mechanism discussed above. The smallest container diameter forms an exception, in the sense that here thick-thin structures are also not found for time scales where they could have been expected (i.e., that lie clearly below the threshold $T$). This behavior maybe due to the fact that lack of material to sustain the collapse leads to an underestimation of the actual times in Eq.~(\ref{timescaleargument2}). But in general the estimate seems to work fairly well.

One remarkable thing is that in our observations a granular eruption (almost) never coincides with a thick-thin structure. This is in agreement with the above mechanism: If an eruption is observed, this means that a relatively large air bubble must have been entrapped. This concurs with a large distance between the first and the second collapse point, which makes it unlikely that a thick-thin structure will be formed. Conversely, if a thick-thin structure is observed, this means that a (relatively small) air bubble must have been pierced by the second jet, which will facilitate its dissolution in the sand.

Incidentally, our observations dismiss the mechanism for the formation of the thick-thin structure put forward in \cite{refRoyerPRE08}, where it was claimed that the rising air bubble pushes up bed material that forms the thick part of the jet. Contrary to the observations, such a mechanism would be stronger for a larger air bubble and cannot explain why a thick-thin structure and a granular eruption cannot be seen at the same time. Next to this --for varying container diameter-- we observe both thick-thin structures and normal jets for the same amount of entrapped air.

Finally, based upon the present experiments we cannot exclude that the second collapse may need some downward motion of the sand bed that would be reinforced in a confined bed, but may become too weak to produce a jet in an unconfined bed. This would mean that if it were possible to increase the container diameter even further, the thick-thin structure may disappear again, as the phase diagrams in Fig.~\ref{thickthin} seem to suggest. This issue needs to be settled in future research.

\section{Conclusions}
\label{conclusions}

In conclusion, we have studied the influence of the boundaries on the different phenomena that can be observed after impact of a ball into a loosely packed sand bed: The penetration of the ball into the bed, the formation of a void, its collapse and the creation of a granular jet, the shape of the granular jet, and the presence of a granular eruption. We have shown that all of the observed behavior is generally well captured by the drag law and hydrostatic collapse model of Section~\ref{Theory}.

More detailedly, in the first part of this study, we have shown that the proximity of the bottom changes these phenomena, starting with the obvious modification of the final position of the ball, which below a certain depth just hits the bottom. The height of the jet is affected, when the void closure is constrained to happen on top of the ball. A granular eruption at the surface only happens if the volume of the entrapped air bubble is large enough, and can be fully suppressed by decreasing the height of the sand bed.

In the second part we have investigated the influence of nearby side walls. Here we find a strong influence on the drag force that the sand bed exerts on the ball when it moves through the sand bed: We find that the hydrostatic drag force component becomes less important, whereas the quadratic (velocity-dependent) component becomes more important. The latter can be traced back to the increased importance of the air flow in the container due to the confinement. Apart from the question why and how the coefficients depend on ambient pressure and container diameter, the drag model of Section~\ref{Theory} provides a quite accurate description of the observations for most of the parameter space. Only the results for the smallest container at low ambient pressure cannot be explained using this framework, due to the constant velocity plateau that is observed during the motion.

The formation and subsequent collapse of the cavity is not only influenced by the modification of the trajectory of the ball; also a smaller amount of sand is involved in its collapse which therefore takes longer for decreased container size. Apart from this, the simple hydrostatic collapse model of a cylindrical cavity presented in Section~\ref{Theory} accounts well for most of the observations. In this way, the modification of the closure time, and closure depth observed in our experiments, can be understood.

As a result of both the changes in the ball's trajectory and the smaller amount of sand that is involved in the collapse, the jet height is affected by the proximity of the wall. In the parameter range of our experiments the unconfined behavior is retrieved when the diameter of the container is larger than $7d$; this value however does seem to depend on the Froude number, and is larger when the Froude number is larger. The occurrence of a granular eruption was shown to be correlated with the size of the air bubble entrapped inside the sand bed.

Finally, this paper culminates in the proposal of a new mechanism for the formation of the thick-thin structure, based upon a second collapse that occurs on top of the ball when it has come to a standstill. To obtain a thick part in the jet, the second jet coming from this secondary collapse needs to be formed fast enough to penetrate the rapidly growing layer of sand that is being created around the point where the first jet had originated.

The work is part of the research program of FOM, which is financially supported by NWO; S. v. K. and S. J. acknowledge financial support.


\begin{thebibliography}{99}
\bibitem{Jaeger1996}
H.M. Jaeger, S.R. Nagel and R.P. Behringer, Rev. mod. Phys., \textbf{68}, 1259-1273 (1996)

\bibitem{refShen01}
S.T. Thoroddsen and A.Q. Shen, Phys. Fluids \textbf{13}, 4 (2001)

\bibitem{refLohsePRL04}
D.Lohse, R. Bergmann, R. Mikkelsen, C. Zeilstra, D. van der Meer, M. Versluis, K. van der Weele, M. van der Hoef, H. Kuipers, Phys. Rev. Lett. \textbf{93}, 198003 (2004)

\bibitem{refRoyerNP05}
J.R.Royer, E.I. Corwin, A. Flior, M.-L. Cordero, M.L. Rivers, P.J. Eng, H.M. Jaeger, Nature Phys. \textbf{1}, 164 (2005)

\bibitem{refRoyerPRL07}
J.R.Royer, E.I. Corwin, P.J. Eng, H.M. Jaeger, Phys. Rev. Lett. \textbf{99}, 038003 (2007)

\bibitem{refRoyerPRE08}
J.R.Royer, E.I. Corwin, B. Conyers, A. Flior, M.L. Rivers, P.J. Eng, H.M. Jaeger, Phys. Rev. E \textbf{78}, 011305 (2008)

\bibitem{refCaballeroPRL07}
G.Caballero, R. Bergmann, D. van der Meer, A. Prosperetti, D. Lohse, Phys. Rev. Lett. \textbf{99}, 018001 (2007)

\bibitem{refMarston08}
J.O. Marston, J.P.K. Seville, Y.-V. Cheun, A. Ingram, S.P. Decent, M.J.H. Simmons, Physics of Fluids \textbf{20}, 023301 (2008)

\bibitem{refUeharaPRL03}
J.S. Uehara, M.A. Ambroso, R.P. Ojha, D.J. Durian, Phys. Rev. Lett. \textbf{90}, 194301 (2003)

\bibitem{refCiamarraPRL04}
M.P. Ciamarra, A.H. Lara, A.T. Lee, D.I. Goldman, I. Vishik, H.L. Swinney, Phys. Rev. Lett. \textbf{92}, 194301 (2004)

\bibitem{refLohseNature04}
D.Lohse, D. Rauh\'e, R.P.H.M. Bergmann, D. van der Meer, Nature (London) \textbf{432}, 689 (2004)

\bibitem{refDeBruyn04}
J. R. de Bruyn and A. Walsh, Can. J. Phys. \textbf{82}, 439 (2004)

\bibitem{refHouPRE05}
M. Hou, Z. Peng, R. Liu, K. Lu, C.K. Chan, Phys. Rev. E \textbf{72}, 062301 (2005)

\bibitem{refTsimring05}
L.S. Tsimring and D. Volfson, in \textit{Powders and Grains 2005}, edited by R. Garcia-Rojo, H.J. Herrmann, and S. McNamara  (Taylor and Francis, London, 2005), p. 1215Ð1223

\bibitem{refKatsuragi07}
Hiroaki Katsuragi and Douglas J. Durian, Nature Physics \textbf{3}, 420-3 (2007)

\bibitem{refDurian2008}
E. L. Nelson, H. Katsuragi, P. Mayor, and D. J. Durian, Phys. Rev. Lett. \textbf{101}, 068001 (2008)

\bibitem{refSequin08}
A. Seguin, Y. Bertho, and P. Gondret, Phys. Rev. E, \textbf{78}, 010301(R) (2008)

\bibitem{refdeVet2007}
S.J. de Vet and J.R. de Bruyn, Phys. Rev. E, \textbf{76}, 041306 (2007)

\bibitem{refMikkelsen02}
R. Mikkelsen, M. Versluis, G.W. Bruggert, E.G.C. Koene, D. van der Meer, K. van der Weele, D. Lohse, Phys. Fluids, \textbf{14}, S14 (2002)

\bibitem{refGilbarg48}
D. Gilbarg and R.A. Anderson, J. Appl. Phys. \textbf{19}, 127 (1948)

\bibitem{refBergman09}
R.P.H.M. Bergmann, D. van der Meer, S. Gekle, A. van der Bos, and D. Lohse, J. Fluid Mech., in press (2009)

\bibitem{footnote1}
Note that the quadratic drag is called "inertial drag" and $\alpha \equiv m/d_0$ where $m$ is the mass of the sphere and $d_0$ is the constant introduced in~\cite{refKatsuragi07}.

\bibitem{footnote2}
Starting from the collapse time $t_c$, the rise time of the bubble has been estimated as that of similarly sized bubble in a liquid experiencing Stokes drag, assuming that it rises in a straight path with its terminal velocity immediately, i.e., from the balance $\phi\rho_g V_{\rm bubble} g \propto \eta V_{\rm bubble}^{1/3} v_{\rm rise}$ (with the packing fraction $\phi$ and the dynamic viscosity $\eta$ assumed to be constant) we have $v_{\rm rise} \propto V_{\rm bubble}^{2/3}g$ with $V_{\rm bubble} \propto (z(t_c) - z_c)d^2$. Now, we estimating the initial position of the top of the bubble as $z_{\rm f} - (z(t_c)-z_c)$. This leads to $t_1 = (z_{\rm f} + z_c - z(t_c))/v_{\rm rise} \propto (z_{\rm f} + z_c - z(t_c))(z(t_c) - z_c)^{-2/3}d^{-4/3}g^{-1}$. The proportionality constant was fitted to give the correct large depth behavior.

Regarding the dissolution time up to $h_{\rm bed} \approx 11\, d$ we can estimate the pressure difference by the hydrostatic pressure in the center of the bubble at the moment the ball has stopped, {\em i.e.} $\Delta P \approx \phi \rho_g g (z_{\rm f} -(1/2)(z(t_c) - z_c))$ (where $z(t_c)$ equals $h_{\rm bed} - d$ for $h_{\rm bed} \leq 5.5\, d$), the volume of entrapped air again as $V_{\rm bubble} \propto (z(t_c) - z_c)d^2$, and, since ball reaches the bottom, the shortest path is around the ball through the sintered plate, i.e., $H \approx d$. Using Darcy's law we have
$t_2 \approx V_{\rm bubble}/Q \propto V_{\rm bubble} H / \Delta P$. Inserting the above quantities we obtain $t_2 \propto (z(t_c) - z_c)d^3 / (\phi \rho_g g (z_{\rm f} -(1/2)(z(t_c) - z_c))$. Above $h_{\rm bed} = 11 \, d$ only the path length changes to $H \propto h_{\rm bed} - z_f$ such that $t_2 \propto (z(t_c) - z_c)(h_{\rm bed} - z_f)d^2 / (\phi \rho_g g (z_{\rm f} -(1/2)(z(t_c) - z_c))$. Again, the proportionality constant was used as a fitting parameter.

\bibitem{footnote3}
The (small) differences in the fitting parameters $\kappa$ and $\alpha$ found for the various Froude numbers were consistent with the measurement error, except for the smallest container diameter at the smallest pressure, as explained in the text.

\bibitem{footnote4}
To obtain Fig.~\ref{eruptionfr100} the closure depths $z_c$ for the diameter of $D=8.5$~cm (which were not measured directly) are obtained by interpolation from Fig.~\ref{collapse_Fr=70}a.

\bibitem{footnote5}
In Fig.~\ref{eruptionfr100} we choose to plot the final depth $z_{\rm f}$ on the vertical axis rather than the ambient pressure $p$ to support the analysis in this Section which makes use of the final depth. Since, as we have shown earlier, $z_{\rm f}$ monotonically decreases with decreasing pressure, the phase diagrams look very similar when using $p$ instead of $z_{\rm f}$ as the quantity plotted on the vertical axis.

\bibitem{footnote6}
Note that the largest container size (the one without an inserted cylinder) has not been included because of its square cross section, which is found to have a marked influence on the jet shape.

\bibitem{footnote7}
At this point it is good to note that such a mechanism explains why the occurrence of a thick-thin structure never seems to interfere with the jet height: The height is determined by the free flight of the thin part which is being formed at the first closure $z_c$, i.e., before the formation of the thick part can become of influence.

\end{thebibliography}
\end{document}